\documentclass[pre,twocolumn,amsmath,amssymb,longbibliography]{revtex4-1}
\usepackage{graphicx,wrapfig,hyperref}
\usepackage[francais]{babel}
\usepackage[latin1]{inputenc}
\usepackage{t1enc}
\usepackage{bm} 
\usepackage{color}
\usepackage{ulem}

\pdfoutput=1

\begin{document}

\title{Burst-and-coast swimmers optimize gait by adapting unique intrinsic cycle\footnote{Significance statement: Body and caudal fin undulations are a widespread locomotion strategy in fish, and their swimming kinematics is usually described by a characteristic frequency and amplitude of the tail-beat oscillation. In some cases, fish use intermittent gaits, where a single frequency is not enough to fully describe their kinematics. Energy efficiency arguments have been invoked in the literature to explain this so-called burst-and-coast regime but well controlled experimental data are scarce. Here we report on an experiment with burst-and-coast swimmers and a numerical model based on the observations to show that the observed burst-and-coast regime can be understood as obeying a minimization of cost of transport.} }


\author{G. Li$^2$, I. Ashraf$^1$, B. Fran\c{c}ois$^1$, D. Kolomenskiy$^2$, F. Lechenault$^3$,
R. Godoy-Diana$^1$, B. Thiria$^1$}

\affiliation{$^1$Laboratoire de Physique et M\'ecanique des Milieux H\'et\'erog\`enes (PMMH), CNRS UMR 7636, ESPCI Paris---PSL University, Sorbonne Universit\'e, Universit\'e de Paris, 75005 Paris, France. \\$^2$Japan Agency for Marine-Earth Science and Technology (JAMSTEC), Yokohama, Japan. \\$^3$Laboratoire de Physique de l'\'Ecole Normale Sup\'erieure (LPENS), 75005 Paris, France}

\begin{abstract}
This paper addresses the physical mechanism of intermittent swimming by considering the burst-and-coast regime of fish swimming at different speeds. The burst-and-coast regime consists of a cycle with two successive phases: a phase of active undulation powered by the fish muscles followed by a passive gliding phase. Observations of real fish whose swimming gait is forced in a water flume from low to high speed regimes are performed, using a full description of the fish kinematics and mechanics. We first show that fish modulate a unique intrinsic cycle to sustain the demanded speed by modifying the bursting to coasting ratio while maintaining the duration of the cycle constant. Secondly, we show using numerical simulations that the chosen kinematics correspond to optimized gaits over the range of swimming speeds tested. 
\end{abstract}

\maketitle

\section{Introduction}

Intermittent dynamics have been frequently observed in fish locomotion. Known as burst and coast, the gait consists of a two-step sequence: an active phase during which fish produce the propulsive force, followed by an inertial, passive phase where they glide or coast without muscular action. This behavior is observed either permanently as part of the strategy of an animal to move and explore its environnement, or during short periods as part of high-speed swimming regimes.  Burst and coast has been addressed extensively by the biomechanics community in the past decades \cite{Weihs:1974,Videler:1982,Blake:1983,Fish:1991,Drucker:1996,Dutil:2007,Tudorache:2007,Paoletti:2014,Calovi:2018,Akoz:2018}, often associated with locomotion cost optimization. Starting from the early studies of Weihs \cite{Weihs:1974}, these works have essentially investigated the relationship between the construction of the burst-and-coast cycle and the global swimming efficiency, when compared to continuous undulatory mechanisms.

Intermittent swimmers minimize the energetic cost of swimming in the gliding phase, during which the fish body is passive and straight, hence not producing mechanical effort and dissipating less into the fluid. The energetically-optimal working point at a given speed is then obtained by tuning the typical times spent in the burst and coast phases of the cycle, balancing the advantage of the passive coasting phase and the energetic injection of the bursting phase to sustain the desired average speed. Most studies have addressed this mechanism theoretically, reducing the problem of burst-and-coast swimming to the optimization of a mechanical system, decoupled from physiological behavior \cite{Videler:1982,Akoz:2018}. Overall, there is a strong lack of experimental data and parametric studies concerning intermittent swimmers, making the question of gait adaption in real animals an open question. 

The purpose of this paper is to examine burst-and-coast swimming using an experimental work performed on live fish that swim using body and caudal fin (BCF) propulsion. A typical burst-and-coast swimmer, the red-nose tetrafish \textit{Hemigrammus bleheri} \cite{Ashraf:2016,Ashraf:2017}, is forced to swim in a flume at a given velocity $U$, and video recordings are used to examine changes in the gait as the imposed velocity $U$ changes. We show that, instead of modulating the frequency and amplitude of the kinematics, the fish rather adapt the burst-to-coast ratio keeping the time $T_ {\rm bout}$ of a typical burst-and-coast event within a narrow range between 0.2 and 0.4 s. The burst phase is a sequence of tail beats with nominally constant frequency and amplitude. More importantly, we demonstrate using a 3D numerical model based on the experimentally measured swimming kinematics, that for a given swimming speed, the burst-and-coast cycle chosen by the fish corresponds to a gait minimizing the global cost of transport. 

\begin{figure*}[htb!]
\begin{center}
\phantom{.} \includegraphics[height=0.4 \linewidth]{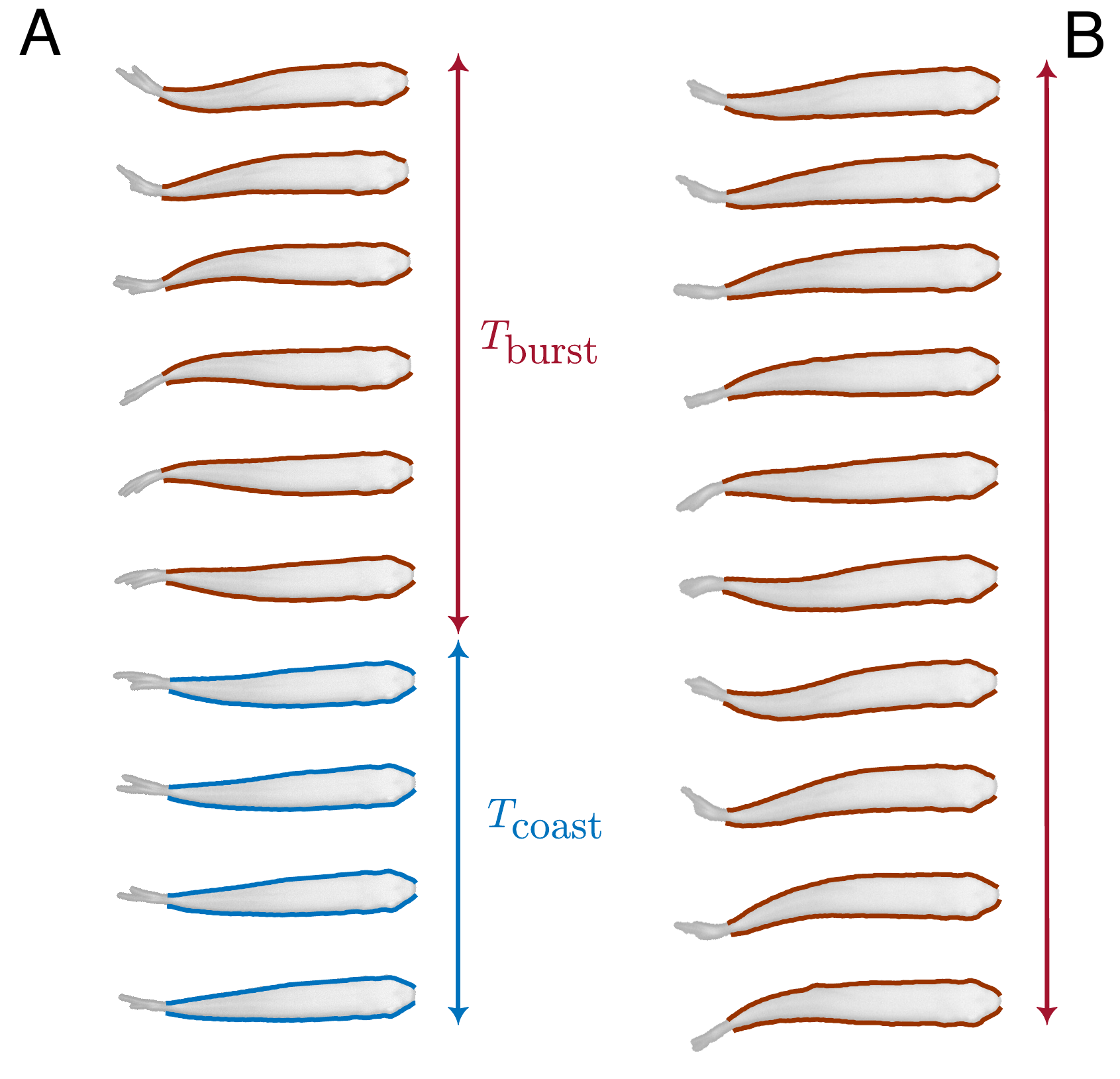}
\includegraphics[height=0.4 \linewidth]{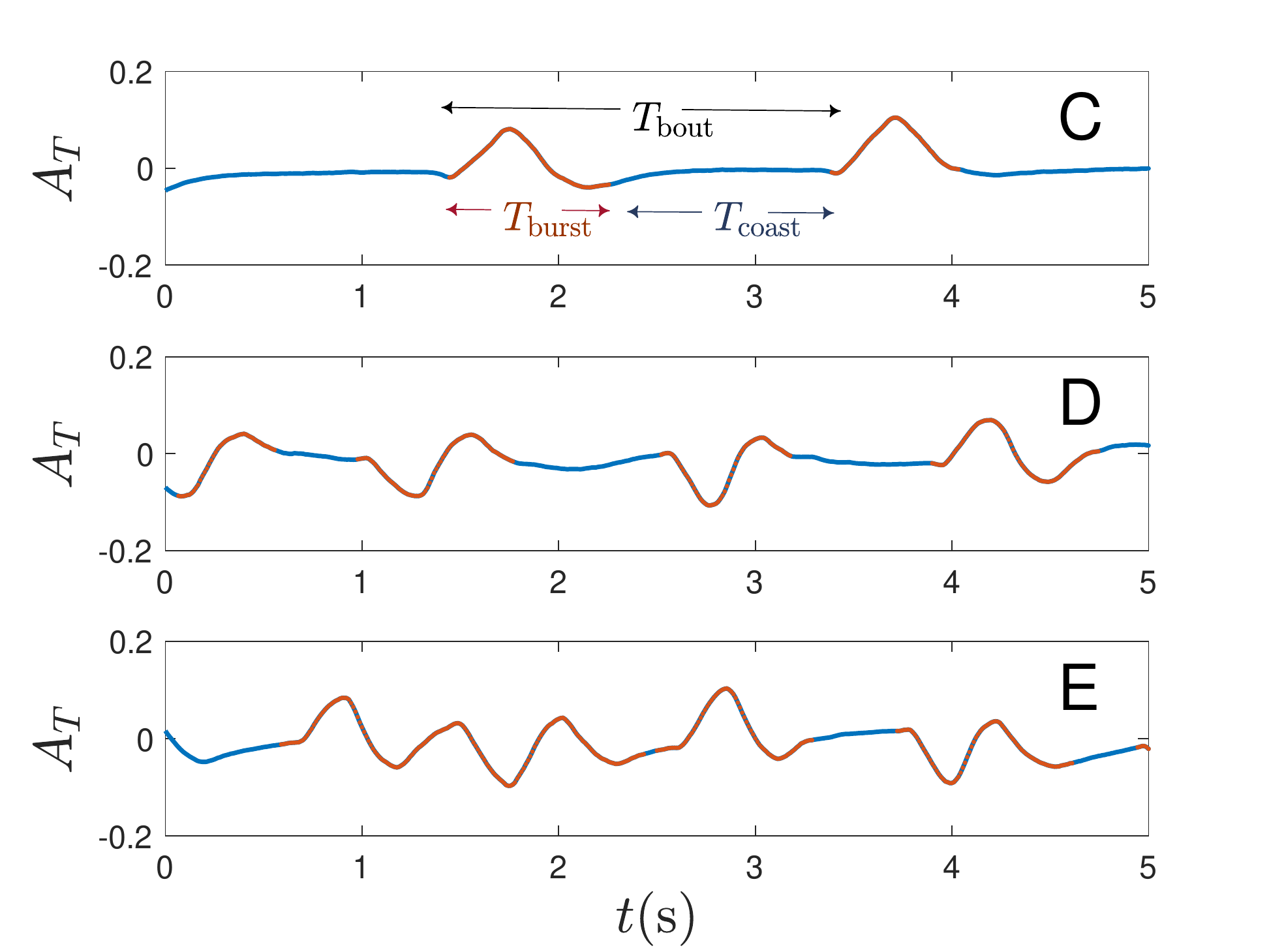}
\caption{(A) and (B) Time steps of typical swimming cycles for two different imposed swimming speeds, (A): $U=1.15$ BL/s and (B) $U=1.9$ BL/s . The swimming cycle is composed of two phases: one \textit{burst} phase where the fish produces propulsion by undulating its body, followed by a \textit{coast} phase during which the fish pauses and glides thanks to its own inertia. As can be seen, the main difference in kinematics between the two gaits (A) and (B) is the time ratio between the propulsive action and the pause. The first part of the swimming cycles are similar, with comparable beating frequency and amplitude of the body undulation but the fish swimming at the larger speed in (B) continues beating for the full cycle while the other pauses before the next one. Colored contours correspond to the fitted fish shape with the elastic model (see text). (C-E) Typical tail-tip kinematics extracted from video analysis for three different imposed swimming velocities. (C): $U=.66$ BL/s, (D): $U=1.15$ BL/s and (E): $U=1.9$ BL/s. As can be observed, the intrinsic characteristics of the active burst cycle (in red color) share the same properties whatever the imposed gait. The frequency and amplitude of the basic tail beat are similar and imposed velocity is sustained just by increasing linearly the burst time with respect to the pause time. }
\label{Fig1}
\end{center}
\end{figure*}

\section{Results}

The experiments were conducted on four individuals of \textit{Hemigrammus bleheri} fish in a water flume with flow velocity varying from 0 to 3 body lengths per second (BL/s). Each fish is recorded in runs of 10 s using fast camera imaging, and the body undulation kinematics is subsequently characterized by extracting the midline of the fish images for each video recording. Additionally, the fish contours are fitted with an elastic beam model of varying thickness (see Materials and Methods). Although such a beam is a crude model of the fish body, it is a very effective tool to examine the deformation kinematics. For each imposed swimming, the measurements were repeated four times, giving a set of 16 different runs for compiling one data point.
 
\begin{figure*}[htb!]
\begin{center}
\phantom{.} \includegraphics[width=.8 \linewidth]{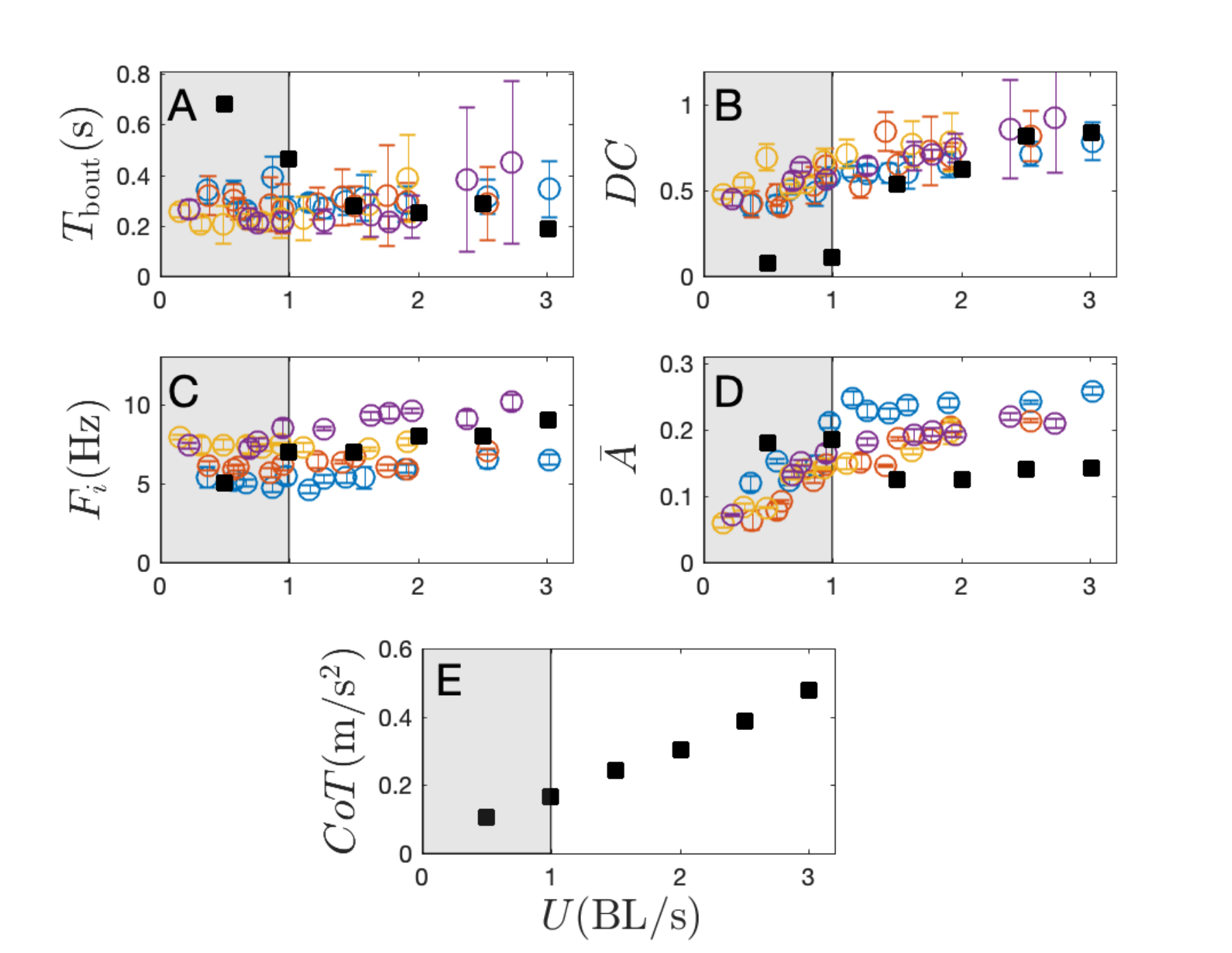}
\caption{Relevant quantities extracted from fish kinematics and beam model fit as a function of the imposed swimming speed $U$ in body length per second (BL/s).  (A): bout duration $T_ {\rm bout}$. (B): Duty cycle $DC=T_{\rm burst}/{T_ {\rm bout}}$. (C) and (D): frequency and peak-to-peak amplitude of the tail beat during a bursting phase, respectively. (E): Cost and transport $CoT$. Different marker colors correspond to different individuals. The optimization results obtained from the burst-and-coast model based on 3D numerical simulations of an artificial swimmer are superimposed to the experimental data in filled black square symbols ($\blacksquare$).}
\label{Fig2}
\end{center}
\end{figure*}

Fig. \ref{Fig1} shows two typical cycles of swimming for a fish at two different imposed velocities: 1.15  BL/s and 1.65 BL/s. The corresponding tail-tip kinematics are also plotted in Fig. \ref{Fig1} (D) and (E). An additional case of lower swimming speed is shown in Fig. \ref{Fig1} (C), to clearly define graphically the burst-and-coast cycle of characteristic time $T_{\rm bout}=T_{\rm burst}+T_{\rm coast}$. A first observation to be made is that the basic undulation cycle of the fish is roughly the same regardless of the swimming speed, i.e. the amplitude and duration of the tail-beat remain visually similar. In order to sustain the increasing swimming velocities between frames (C) to (D) of Fig.~\ref{Fig1}, the strategy seems to be to increase the number of tail beats within the burst, while the coasting time is diminished.
This observation is confirmed by the results obtained for the all different individuals and swimming speeds. Fig. \ref{Fig2} shows all the relevant quantities of the swimming kinematics: the characteristic duration of the full burst and coast swimming cycle  $T_{\rm bout}$ in Fig. \ref{Fig2} (A); the duty cycle, i.e. the time ratio of the burst phase and the full burst-and-coast bout $DC={T_{\rm burst}}/{T_ {\rm bout}}$ (Fig. \ref{Fig2} (B)); and the typical tail beat frequency $F_i$  and scaled amplitude $\bar{A}=A/L$ in the burst phase---Fig. \ref{Fig2} (C) and (D), respectively. 

In addition, the cost of transport $CoT$ obtained from the numerical model is shown as a function of the swimming speed in Fig. \ref{Fig2} (E). The cost of transport is defined as

$$CoT=\frac{\overline{P}}{m\overline{U}}\; ,$$

\noindent i.e. as the power $\overline{P}$ normalized by the average swimming speed $\overline{U}$ and the mass of the fish  (see \cite{Li:2019} and Material \&  Methods for details). 

Fig. \ref{Fig2} clearly shows that fish have two internal times on which they construct the burst-and-coast kinematics to attain the desired gait: $T_{\rm bout}$ and $T_{i}=1/F_i$ . The first internal time, $T_{\rm bout}$, is constrained essentially to a range from 0.2 to 0.4~s for all fish---Fig. \ref{Fig2} (A). Thus, it seems that the full burst-and-coast cycle is independent of the swimming speed: fish do not modulate the time between two phases of action regardless of the imposed velocity.  Other works have already reported the regularity of the burst-and-coast swimming for different species and different stages of maturity (see for instance \cite{Fuiman:1988,Muller:2000,McHenry:2005,Olive:2016,Harpaz:2017}), and this has been attributed to neural sensing mechanisms \cite{Olive:2016,Dunn:2016}. It is also worth noting that this time remains fairly constant across individuals. The second characteristic time $T_{i}=1/F_i$ is the inverse of the internal frequency $F_i$ shown in Fig. \ref{Fig2} (C). The absolute value of $F_{i}$ seems related to each specific fish, but it is statistically constant for all velocities tested in the experiment, meaning that fish do not modulate this internal tail beat frequency as a gait adapting strategy. Concerning the amplitude of the burst cycle, Fig. \ref{Fig2} (D) shows that the magnitude of the tail beat increases in the range of slow velocity (0 BL/s$<U<$1 BL/s) to saturate afterwards at higher speeds. This can be readily understood recalling that the burst is short at low velocities, so that there is no time to accommodate more than one half of a period of the tail-beat oscillation---see the burst profiles in Fig. \ref{Fig1} (C). At last, we see that the duty cycle $DC$  (i.e. the fraction of the burst-and-coast bout during which the fish are actively producing thrust) increases linearly with the demanded swimming speed. 

\section{Discussion}

The reading of Fig. \ref{Fig2} tells that intermittent swimmers repeat an intrinsic basic movement to sustain the desired swimming speed. This movement consists of an active undulation of constant frequency and almost constant tail beat amplitude (except in the low-speed range, shaded in gray in the panels of Fig.~\ref{Fig2}), repeated as long as it is needed. Thus, a fish willing to swim twice as fast will double its bursting time. Of course, because each burst-and-coast swimming sequence is performed over a constant time $T_{\rm bout}$, fish spending more time in the burst phase necessarily also shorten the coast duration, which sets an upper limit to the swimming speed that can be achieved.  It is interesting to note that the swimming behavior described here differs from the idea that fish modulate their body wave kinematic parameters to change speed, in contrast with what has been observed for larger fish using continuous swimming---see for instance \cite{Lowe:1996,Webb:1984}. To our knowledge, such a mechanism has not been reported in the literature, especially concerning small-sized fish of a few centimetres as the tetra fish of the present experiments.

In order to understand the dynamics underlying the experimental observations, we studied the swimming optimization problem of a simulated burst-and-coast swimmer. The fish is modeled using the realistic body geometry of \textit{Hemigrammus bleheri} extracted from the experiment (see Supplementary Information). The burst-and coast cycle is built, following the observations, by concatenating an active phase and a passive phase. The flow field around the fish during each phase is simulated using computational fluid dynamics (CFD)---see Materials and Methods. The free parameters used in the swimming simulation are: the duty cycle $DC$, the tail beat duration $T_i$ (or internal frequency $F_i$)  and the tail beat amplitude $\bar A$. For each swimming velocity, the set of parameters ($DC$, $F_i$, $\bar A$, $T_{\rm bout}$) that minimizes the cost of transport ($CoT$) is selected. The results of the optimization procedure are superimposed to the experimental data in Fig. \ref{Fig2} (black squares). 

For moderate to high speeds (in the range 1 BL/s $<U<$ 3 BL/s), the parameters that minimize the energy cost of swimming match closely the experimental data. This is a remarkable observation, as it shows that fish in the range of cruise swimming speeds constantly optimize their $CoT$. Moreover, such optimization mainly consists in maintaining the tail-beat frequency and amplitude constant and modulating the time of bursting. However, the predictions of the optimization procedure fail to reproduce the observations in the low-speed range (0 BL/s $<U<$ 1 BL/s). The optimization predicts a larger $T_{\rm bout}$ and a smaller $CoT$. The fact that fish do not increase $T_{\rm bout}$ at low swimming speeds may have different possible explanations---such as the physiological constraints of muscle efficiency or the sensorimotor capacity necessary for maintaining the body orientation---, but it may also be explained considering that $CoT$ minimization might not be needed at such low swimming velocities due to the low energy consumption involved.

The remarkable agreement between the optimization calculation and experimental observations leads us to two important conclusions for burst and coast swimmers: 1) fish essentially do not modulate tail-beat frequency as observed for continuous swimming \cite{Lowe:1996,Webb:1984} but adapt a unique cycle to sustain the imposed speed; and 2) the frequency, amplitude of the tail beat and the burst phase duration (the duty cycle) are optimal parameters with respect to the cost of transport $CoT$ at typical cruise speeds. It is also noteworthy that the results of the simulation are not exclusively associated to the species \textit{Hemigrammus bleheri}. Excepted the details of the body shape that were extracted from the experiments, the construction of the intermittent simulated kinematics (see Materials and Methods) uses a generic body deformation that can describe other burst-and-coast swimmers. The results presented in this paper bring thus a general description of intermittent fish locomotion, based on experimental observations: because of the intermittency constraint---the bout time, most likely fixed because of physiological reasons---, these fish have developed specific swimming sequences minimizing their cost of transport that are different form those observed for continuous swimmers. Future works should multiply experimental observations and produce a larger inventory of intermittent swimmers to determine if the burst-and-coast mechanism described here holds for other fish species. 

\begin{figure*}[tb!]
	\begin{center}
		\phantom{.} \includegraphics[width=0.65 \linewidth]{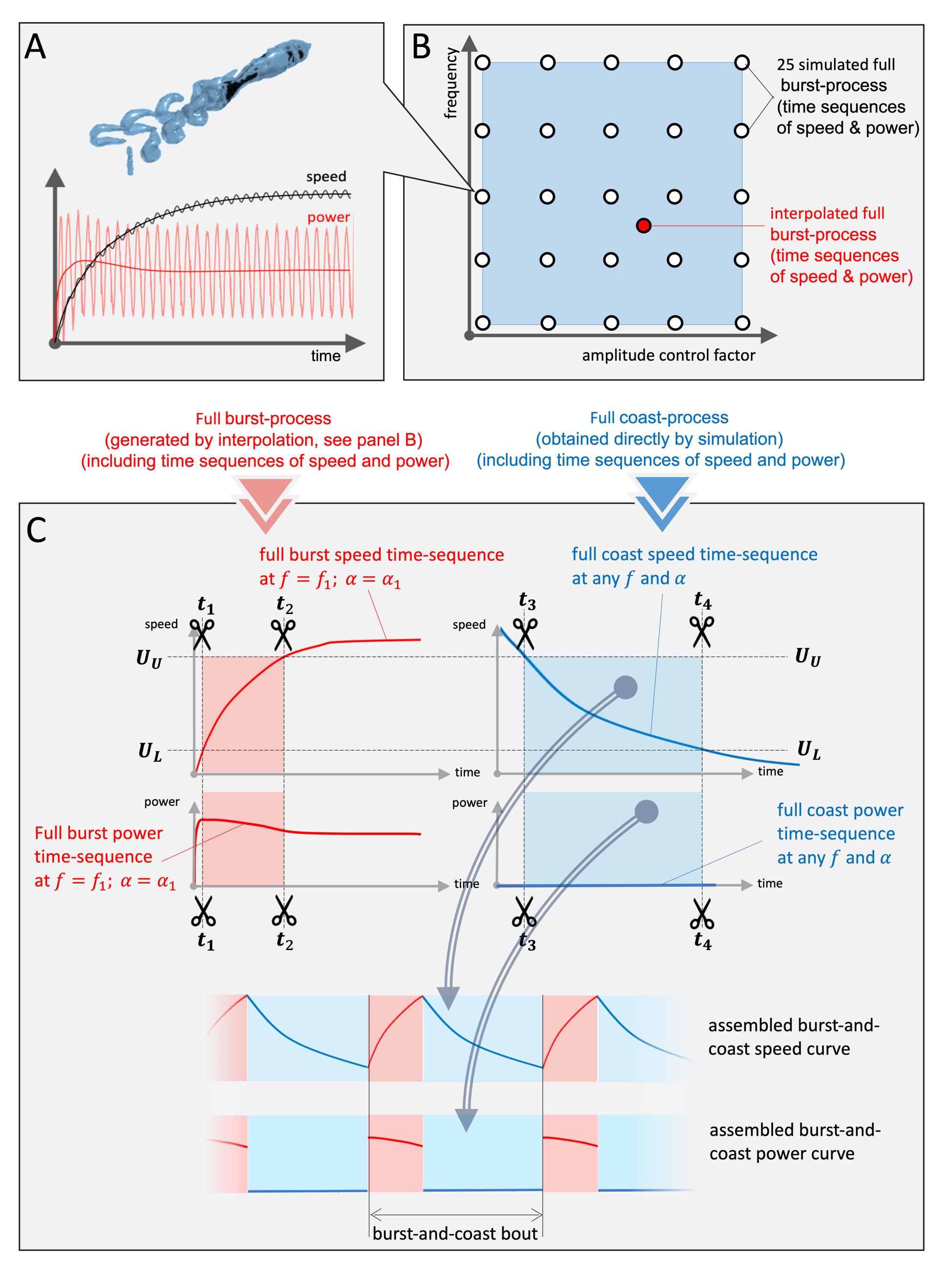}
		\caption{Graphical representation of the numerical burst-and-coast modeling. (A): Simulation of a full burst-process based on a three-dimensional self-propelled fish swimming model. In each simulation, the fish accelerate from a static condition until a stable speed is reached. The time sequences of speed and power are recorded. (B): Based on 25 sample simulations, for an arbitrary combination of tail beat frequency and amplitude, the corresponding full burst-process (time sequences of speed and power) is obtained by interpolation. (C): Full burst-process and full-coast process are trimmed according to and assembled into the burst-and-coast gait.}
		\label{Fig:model}
	\end{center}
\end{figure*}

\section{Materials and Methods}

\subsection{Animals and housing} Red nose tetra fish \textit{Hemigrammus bleheri} of body length in the range $\sim$3.5-4 cm long and height $\sim$0.5-0.6 cm, were procured from a local aquarium supplier (anthias.fr, France). The fish were reared in a 60 litres aquarium tank with water at a temperature between 26-27 $^{\circ}$C and they were fed 5-6 times a week with commercial flake food. Results from experiments with four individuals are analyzed here, for which a full set of different swimming speeds were recorded. The experiments performed in this study were conducted under the authorization of the Buffon Ethical Committee (registered to the French National Ethical Committee for Animal Experiments no. 40).

\subsection{Swimming flume} A shallow water tunnel with a test section of 2.2cm depth and a swimming area of 20cm $\times$ 15cm was used for the experiments---see also \cite{Ashraf:2016,Ashraf:2017}, where the same setup has been used to study collective swimming dynamics. The flow rate $Q$ can be varied from 4 to 22 litres per minute, resulting in an average velocity $U=Q/S$, where $S$ is the cross section, in the range between 2.7 cm.s$^{-1}$  to 15 cm.s$^{-1}$. 
The mean turbulence intensity in the channel is between 3-5\% (characterized using PIV in previous work \cite{Ashraf:2017}) and is independent of the flow rate. The velocity profile in the mid-section of the channel is rather flat and also remains unchanged for the different flow rates used, the wall effect region being limited to a distance smaller than 3mm.

\subsection{Experimental procedure}
Before each measurement, the fish group was transferred to the swimming tunnel with the fluid at rest and left for around one hour in order to acclimatize to the conditions of the experiments. The swimming runs were carried out for 10 seconds on each individual, increasing gradually the imposed speed from 0.5 BL/s to 3 BL/s. The procedure was then repeated several times, with a typical 30-minute resting pause between measurements. The complete set of experiments consisted in 3 to 5 runs per individual, at 10 different velocities, on a group of 4 individuals, corresponding to 150 measurements points.

\subsection{Data statistics} 
The tail-beating kinematics was extracted for each fish in a group. The average and standard deviation were computed for each group and each velocity. All points presented in Fig. \ref{Fig2} are thus averaged quantities with several experiments per point. For instance, the data points for the frequency are given by  $F_i=\frac{1}{N}\sum_{0}^{N} {\langle f\rangle}$ over the number of individuals and the different runs, where the brackets denote a time average and $N$ is the number of individuals within the school. Thus, we obtain a single value of the frequency for each group size and swimming speed.



\subsection{Burst-and-coast numerical model}
We developed a numerical model that can generate an arbitrary burst-and-coast swimming gait in a four-dimensional parameter space. The parameters are: 1) the frequency of the burst phase $f_b$, 2) the amplitude of the burst phase $A_b$, 3), the upper speed bound $U_U$ (the speed at which the fish stops bursting and starts coasting) and 4) and lower speed bound $U_L$ (the speed at which fish stops coasting and starts bursting, $U_L<U_U$). Then, we search across this parameter space for an optimal burst-and-coast swimming gait that guarantees sustained swimming with some specified speed $\overline{U}$ at the lowest cost of transport $CoT$. The numerical solutions of this constrained optimization problem involve a coarse discretization of the parameter space, a composition of a data base of different gaits with those few discrete values of the frequency and amplitude, and a subsequent interpolation using that data base.

The data for the burst phase are obtained my means of computational fluid dynamics (CFD) simulations using a well-validated three-dimensional solver based on the overset-grid finite-volume method (\cite{Liu:2009,Li:2012}; for more information including the numerical validation, see Supplementary Information. We simulated ``full burst processes" of a self-propelled fish in continuous swimming with some constant frequency and amplitude. The bodylength of the modle fish is 2cm and its deformation is driven by a sinusoidal function. The fish accelerated from rest in quiescent water until it nominally reached its maximum speed---figure \ref{Fig:model} (A). Since such a ``full burst process" is fully determined by the tail beat frequency and amplitude of the fish, we simulated 25 cases with 5 different frequencies ($2$, $6$, $10$, $14$ and $18$Hz) and 5 different tail beat amplitudes (approximately $0.02$, $0.07$, $0.13$, $0.19$, and $0.26L$). The range of the Reynolds number in this study is below $6000$, turbulence models are not used, and the grid resolution at $Re=6000$ has been justified in a previous study (\cite{Li:2014}). The data from all full burst process cases were low-pass filtered to remove the periodic fluctuation caused by the tail beat. Using these 20 cases as interpolation nodes, one can quantify any arbitrary full burst process with some specified tail beat amplitude and frequency---see figure \ref{Fig:model} (B). 

The coast phase motion and energetics data were obtained using the same CFD solver mentioned above, letting the model fish stop undulating after reaching the speed of $13$BL/s (the highest speed reached across all simulated cases corresponding to $f_b=18Hz$ and $A_b = 0.26L$). During this coast phase, the body was held straight and the fish decelerated until the velocity dropped to almost zero. Note that the mechanical power consumption in the coast phase is zero.

Thus, a burst-and-coast process is defined when an upper speed bound $U_U$ and a lower speed bound $U_L$ are specified. The full burst and the full coast data sequences are trimmed according to the values of $U_U$ and $U_L$, respectively. The full swimming cycle was obtained by concatenating the trimmed burst and coast time sequences considering that the transition between the burst and the coast phases is instantaneous. The procedure is then duplicated to produce a sawtooth-wave time profile of the velocity---see figure \ref{Fig:model} (C). 

For a given set of the four parameters ($f_b$, $A_b$, $U_U$ and $U_L$), as long as $U_U$ and $U_L$ are within the speed range of the ``full burst process", we obtain a unique burst-and-coast swimming gait. The average speed of the generated burst-and-coast swimming gait is defined as $\overline{U}$, the average power as $\overline{P}$ and the CoT as $\overline{P}/\overline{U}$. We programmed a MATLAB code to scan the four parameter dimensions  in order to find an optimal burst-and-coast swimming gait that would meet the required speeds with the lowest cost of transport ($f_b$, scan resolution 1Hz, range $2 \sim 18$Hz; $A_b$ scan resolution approximately $0.015L$, range approximately $0.02 \sim 0.26L$; the scan resolution in $U_U$ and $U_L$ is less than ${10}^{-6}$ $L$/s).

For further details of the numerical model, see Supplementary Information.

\bigskip
\noindent\textbf{Acknowledgements}
{We acknowledge Prof. Hao Liu for his valuable contribution on the CFD model; the personnel from the PMMH Laboratory workshop, in particular Xavier Benoit-Gonin, for their technical help in the construction of the swimming channel; and Jos\'e Halloy for his help with the fish handling protocol.}





\end{document}